\documentstyle[manuscript,aps]{revtex}
\begin{document}
\preprint{
\begin{tabular}{r}
MIT-CTP-2571 \\
hep-ph/9609476\\
Revised Dec. 96
\end{tabular}
}
\begin{titlepage}
\title{Flavor Oscillations in Field Theory}
\author{Elisabetta Sassaroli \thanks{This work is supported in part by funds 
provided by the U.S. Department of Energy (D.~O.~E.) under 
cooperative research agreement $\#$DF-FC02-94ER40818 and in 
part by the University and INFN of Perugia.}  }
\address{Laboratory for Nuclear Science and Department of Physics}
\address{Massachusetts Institute of Technology}
\address{Cambridge, MA 02139}
\address{Dipartimento di Fisica and sezione INFN} 
\address{Universita\' \  di Perugia} 
\address {I-06123 Perugia, Italy}
\maketitle

\begin{abstract}
Neutrino flavor oscillations are discussed in terms of two coupled
Dirac fields. The interacting  Lagrangian is diagonalized to 
obtain the exact eigenvalues and eigenfunctions. Flavor wave functions are 
then derived  directly from the quantized  neutrino fields. Probability
densities obtained by squaring these wave functions upon 
taking into account the 
neutrino chirality are in agreement with the standard neutrino oscillation 
probabilities. 
\end {abstract}

\thispagestyle{empty}

\end{titlepage}

\section{Introduction}

The theory and phenomenology of neutrino flavor mixing has been extensively
studied mainly in the framework of quantum mechanics \cite{Bil,Boe,Bah}. 
Only very
recently, a quantum field theoretical analysis of flavor mixing has been
considered using the LSZ formalism \cite{Blas}.

Given the conflicting experimental results which have been obtained from a
variety of neutrino observations, it is important to investigate the problem
of neutrino propagation in the general context of field theory to reassure
ourselves that there are indeed no differences between these results and
standard ones obtained from the quantum mechanical treatment. Otherwise
said, we need to firmly establish the approximations necessary to derive the
expressions used phenomenologically. It is perhaps not superfluous to point
out that in the Dirac theory, the contribution of negative energy states
becomes substantial, an aspect of the problem totally absent in the quantum
mechanical treatment.

Our discussion in the present paper will be through the following Lagrangian%
$$
{\cal L}={\bar \psi }_e(i\gamma \cdot \partial -m_e)\psi _e+{\bar \psi }_\mu
(i\gamma \cdot \partial -m_\mu )\psi _\mu -\delta ({\bar \psi }_e\psi _\mu +{%
\bar \psi }_\mu \psi _e),\eqno(1.1) 
$$
consisting of two coupled Dirac neutrino fields $\psi _e$, $\psi _\mu $ with
masses $m_e$ and $m_\mu $. The interaction is provided through a lepton
number violating term with a coupling constant $\delta $. The model allows
for exact diagonalization. Neutrino and anti-neutrino flavor wave functions
can be directly obtained as matrix elements of the quantized neutrino fields.

The fully interacting Lagrangian should also include the weak interaction
term. It is ignored here (as is done usually), since we are interested in
free propagation and possible oscillation in flavor alone. For its
production, we simply assume that the neutrino is created through some weak
interaction process. Such a breakup is of course an approximation. The
effect of parity non-conservation due to the weak interaction term on the
other hand, is taken into account by considering the left-handed
(right-handed) components of the neutrino (anti-neutrino) flavor wave
functions.

The model discussed here represents an example (albeit simple) of an
interacting field theory which is exactly solvable. This may be of some help
in elucidating the properties of interacting theories, which can be normally
studied only in perturbation theory. For example, the interaction between
the electron and muon neutrino fields produces a change in their masses: the
``experimental'' masses $m_1$, $m_2$ depend on the ``bare'' muon and
electron neutrino masses $m_e$, $m_\mu $ and the coupling constant $\delta $.

The paper is organized as follows. In Sec. 2, the standard neutrino
oscillation phenomenology is reviewed. In Sec. 3, the two coupled Dirac
equations, obtained from the Lagrangian (1.1) are solved. The electron and
muon neutrino fields are then quantized in terms of the two free uncoupled
fields, which diagonalize Eq. (1.1), as described in Sec. 4. The total
conserved charge is the sum of the electron and muon flavor charges, which
are not conserved separately. The last section closes with some concluding
remarks.

\section{Standard Treatment}

We will critically review the standard quantum mechanical treatment of
neutrino oscillations to bring out the essential approximations made
implicitly and the boundary conditions which are imposed.

A state vector $|\psi >$ is introduced as a linear combination of the flavor
eigenstates $|e>$ and $|\mu >$ (assuming just two flavors) 
$$
|\psi >=C_e|e>+C_\mu |\mu >,\eqno(2.1a) 
$$
$$
|\psi >=\left( \matrix { C_e \cr 
       C_{\mu} \cr }\right) ,\eqno(2.1b) 
$$
with $C_e=<e|\psi >$, $C_\mu =<\mu |\psi >$ and 
$$
|C_e|^2+|C_\mu |^2=1.\eqno(2.1c) 
$$
$C_e$ and $C_\mu $ then become the amplitudes for detecting an electron
neutrino and a muon neutrino respectively.

To derive the time evolution of the coefficients $C_e(t)$ and $C_\mu (t)$,
the state vector $|\psi >$ is written as a superposition of the energy
(mass) eigenstates $|\nu _1>$ and $|\nu _2>$ 
$$
|\psi >=C_1|\nu _1>+C_2|\nu _2>,\eqno(2.2a) 
$$
$$
|\psi >=\left( \matrix { C_1 \cr 
       C_2 \cr }\right) ,\eqno(2.2b) 
$$
with $C_1=<\nu _1|\psi >$, $C_2=<\nu _2|\psi >$ and 
$$
|C_1|^2+|C_2|^2=1,\eqno(2.2c) 
$$
where $C_1$ and $C_2$ are the amplitudes for finding the neutrino in the
energy states $E_1$ and $E_2$ respectively. These coefficients evolve in
time as 
$$
C_1(t)=C_1(0)e^{-iE_1t},\ \ \ \ C_2(t)=C_2(0)e^{-iE_2t}.\eqno(2.3) 
$$

Introducing the rotation matrix between flavor and mass eigenstates 
$$
\left( \matrix {|\nu_1> \cr 
 |\nu_2> \cr }\right) =\left( 
\matrix {cos(\theta) &  -sin(\theta) \cr 
  sin(\theta) & cos(\theta) \cr }\right) \left( \matrix {|e> \cr 
|\mu>}\right) ,\eqno(2.4) 
$$
it is easy to see that the following relation between the energy and flavor
amplitudes holds 
$$
\left( \matrix {C_1(t) \cr 
C_2(t)\cr }\right) =\left( 
\matrix {cos(\theta) 
& -sin(\theta) \cr
sin(\theta) & cos(\theta) \cr }\right) \left( 
\matrix { C_e(t) \cr
C_{\mu}(t)}\right) .\eqno(2.5) 
$$

Hence, the time evolution of the coefficients $C_e$ and $C_\mu $ is given by 
$$
\left( \matrix {C_e(t) \cr 
C_{\mu}(t)\cr }\right) =\left( 
\matrix {cos(\theta) 
& sin(\theta) \cr
-sin(\theta) & cos(\theta) \cr }\right) \left( 
\matrix { C_1(0)e^{-iE_1t} \cr
C_2(0)e^{-iE_2t}}\right) .\eqno(2.6) 
$$

In Eq. (2.6), the boundary condition that at production we have only a given
flavor must be imposed. Suppose for example that at $t=0$ a muon neutrino is
produced, i.e. 
$$
C_\mu (0)=1,\ \ \ C_e(0)=0.\eqno(2.7) 
$$
>From Eq. (2.5) at $t=0$ we obtain%
$$
C_1(0)=-sin(\theta ),\ \ \ \ C_2(0)=cos(\theta ).\eqno(2.8) 
$$

The time evolution of the flavor amplitudes is obtained by substituting Eq.
(2.8) in Eq. (2.6) 
$$
C_e(t)=sin(\theta )cos(\theta )(e^{-iE_2t}-e^{-iE_1t}),\eqno(2.9a) 
$$
$$
C_\mu (t)=sin^2(\theta )e^{-iE_1t}+cos^2(\theta )e^{-iE_2t}.\eqno(2.9b) 
$$

Space and therefore momentum is introduced by assuming in Eqs. (2.9) $%
E_1^2=m_1^2+p^2$, $E_2^2=m_2^2+p^2$ and $x= t$. The probability of
finding a given flavor is obtained by squaring Eqs. (2.9) 
$$
|C_e|^2=sin^2(2\theta )sin^2[(E_2-E_1)t/2]\simeq sin^2(2\theta )sin^2\left[ {%
\frac{(m_1^2-m_2^2)x}{2p}}\right] ,\eqno(2.10a) 
$$
$$
|C_\mu |^2=1-sin^2(2\theta )sin^2[(E_2-E_1)t/2]\simeq 1-sin^2(2\theta
)sin^2\left[ {\frac{(m_1^2-m_2^2)x}{2p}}\right] .\eqno(2.10b) 
$$

The assumption that the muon neutrino is created with a definite momentum $p$
is only an approximation as has been pointed out previously
\cite{Win,Kay,Giun,Sriv}. It is in contradiction 
with four momentum conservation,
for example for the reaction $\pi \rightarrow \mu \nu $. Each of the
possible energy eigenstates has a somewhat different momentum ${\bf p}_i$.
In the rest frame of the pion, energy conservation dictates that (i= 1, 2) 
$$
M_\pi =\sqrt{M_\mu ^2+{\bf p}_i^2}+\sqrt{m_i^2+{\bf p}_i^2}.\eqno(2.11) 
$$

Therefore, if we introduce momentum, i.e. space in Eqs. (2.9) we should
write 
$$
C_e(x,t)=sin(\theta )cos(\theta )(e^{ip_1x}e^{-iE_1t}-e^{ip_2x}e^{-iE_2t}),%
\eqno(2.12a) 
$$
$$
C_\mu (x,t)=sin^2(\theta )e^{ip_1x}e^{-iE_1t}+cos^2(\theta
)e^{ip_2x}e^{-iE_2t}.\eqno(2.12b) 
$$
In the relativistic approximation $x\simeq t$ the squared moduli of the
amplitudes $C_e(x,t)$, $C_\mu (x,t)$ reduce to the standard neutrino
oscillation probabilities given by Eqs. (2.10), as described 
in Ref. \cite{Win}.
The semiclassical approximation that a neutrino moves  at 
a velocity close to $c$ (assuming very small neutrino masses) ``on a classical 
path'' ($x=ct$) can  be a good
approximation if the neutrino travels over a macroscopic distance.

Hence, if a muon neutrino is produced at the space-time $(x\simeq ct=0)$,
the probability of observing an electron neutrino is maximum for those
space-time points $x\simeq ct$ at which $|C_e|^2=1$. If some matter is
present at these points, processes such as $\nu _e+n\rightarrow p+e^{-},$
will occur, but not processes of the type $\nu _\mu +n\rightarrow p+\mu ^{-}$
indicating therefore flavor oscillations.

\section{Field Theoretical Discussion}

Since neutrinos are relativistic particles of spin 1/2, it is important to
derive a relativistic equation of motion which can describe such flavor
mixing. Energy eigenfunctions can be derived from this equation and one
proper way to deal with states of negative energies is to quantize the
field. As stated in the Introduction, we will  consider the following
interacting Lagrangian 
$$
{\cal L}={\bar \psi }_e(i\gamma \cdot \partial -m_e)\psi _e+{\bar \psi }_\mu
(i\gamma \cdot \partial -m_\mu )\psi _\mu -\delta ({\bar \psi }_e\psi _\mu +{%
\bar \psi }_\mu \psi _e).\eqno(3.1) 
$$
The parameter $\delta $ is an extra mass (energy) related to the small 
amplitude
that a neutrino can flip flavor. The following two coupled Dirac equations 
$$
(i\gamma \cdot \partial -m_e)\psi _e-\delta \psi _\mu =0,\eqno(3.2) 
$$
$$
(i\gamma \cdot \partial -m_\mu )\psi _\mu -\delta \psi _e=0,\eqno(3.3) 
$$
describe a neutrino which has some probability to flip flavor. If no flavor
flips were possible ($\delta =0$), the rest energies (masses) of the system
would be $m_e$ and $m_\mu $, possibly equal to zero. However, since there is
some amplitude that a neutrino, which is produced as an electron 
neutrino becomes later a muon neutrino, the possible rest 
energies of the system are
not simply $m_e$ and $m_\mu $, but are functions of the flipping energy.

It is easy to see that the conserved current is 
$$
J^\mu ={\bar \psi }_\nu \gamma ^\mu \psi _\nu ={\bar \psi _e}\gamma ^\mu
\psi _e+{\bar \psi _\mu }\gamma ^\mu \psi _\mu =J_e^\mu +J_\mu ^\mu .%
\eqno(3.4) 
$$
Thus, the separate electron and muon flavor currents are not conserved, only
their sum is.

In order to determine the energy eigenvalues and eigenfunctions of the
system of equations (3.2) and (3.3)  we consider the ansatz 
$$
\psi _e=ae^{-iPx},\eqno(3.5) 
$$
$$
\psi _\mu =be^{-iPx},\eqno(3.6) 
$$
where $P$ is the four-momentum $P=(E,{\bf p})$, which is unknown and is to
be determined so that the system of differential equations (3.5) and (3.6)
is satisfied. The coefficients $a$ and $b$ are Dirac spinors, which can be
written as 
$$
a=\left( \matrix {\chi_1 \cr
		    \chi_2 \cr}\right) ,\eqno(3.7a) 
$$
$$
b=\left( \matrix {\varphi_1 \cr
		    \varphi_2 \cr}\right) ,\eqno(3.7b) 
$$
where $\chi _{1,2}$ and ${\varphi _{1,2}}$ are two component vectors.
Substituting Eqs. (3.5), (3.6) into Eqs. (3.2), (3.3), we obtain the system
of linear homogeneous equations 
$$
E\chi _1={\bf \sigma }\cdot {\bf p}\chi _2+m_e\chi _1+\delta \varphi _1,\ \
\ \ E\chi _2={\bf \sigma }\cdot {\bf p}\chi _1-m_e\chi _2-\delta \varphi _2,%
\eqno(3.8a) 
$$
$$
E\varphi _1={\bf \sigma }\cdot {\bf p}\varphi _2+m_\mu \varphi _1+\delta
\chi _1,\ \ \ E\varphi _2={\bf \sigma }\cdot {\bf p}\varphi _1-m_\mu \varphi
_2-\delta \chi _2,\eqno(3.8b) 
$$
where $\sigma $ are the Pauli matrices.

The system of Eqs. (3.8) admits non trivial solutions only if 
$$
E^4-E^2(2p^2+2\delta ^2+m_e^2+m_\mu ^2)+p^4+\delta ^4+ 
$$
$$
p^2(2\delta ^2+m_e^2+m_\mu ^2)+m_\mu ^2m_e^2-2\delta ^2m_em_\mu =0.\eqno(3.9)
$$
Solving Eq. (3.9), we obtain ($p=|{\bf p}|$) 
$$
E_{1,2}=\pm \sqrt{p^2+m_{1,2}^2},\eqno(3.10) 
$$
with $m_{1,2}$ given by 
$$
m_{1,2}={\frac 12}[(m_e+m_\mu )\pm R],\eqno(3.11) 
$$
and 
$$
R=\sqrt{(m_\mu -m_e)^2+4\delta ^2}.\eqno(3.12) 
$$

Therefore, while in the free Dirac equation there are two energies (one
positive and one negative) for every possible value of the momentum ${\bf p}$, 
for a system of two coupled Dirac equations, for every possible value of
the momentum, there are four possible values of the energy, two positive and
two negative, due to the possibility of (flavor) oscillations. Also, because
there is some chance that the neutrino can flip flavor, the rest
energies of the electron and muon neutrino system are not simply $m_e$, $%
m_\mu $ but are given by Eq. (3.11).

The solutions of Eqs. (3.8) can be written as 
$$
\chi _1=-{\frac{\delta ({\bf \sigma }\cdot {\bf p})(m_e+m_\mu )}Q}\varphi
_2,\ \ \ \chi _2={\frac{\delta (E^2-p^2-\delta ^2-E(m_e+m_\mu )+m_em_\mu )}Q}%
\varphi _2, 
$$
$$
\varphi _1={\frac{(\sigma \cdot {\bf p})(p^2+m_e^2+\delta ^2-E^2)}Q}\varphi
_2,\eqno(3.13) 
$$
with 
$$
Q=-E^3+E^2m_\mu +E(\delta ^2+p^2+m_e^2)+\delta ^2m_e-p^2m_\mu -m_e^2m_\mu .%
\eqno(3.14) 
$$
\smallskip
As stated above, for a given value of the momentum ${\bf p}$, there are four
different energies $\pm E_{1,2}$ and for each energy, there are two
eigenfunctions with different (up and down) spins.

Corresponding to the positive energy solution $E_1=\sqrt{m_1^2+p^2}$, we
have the following two solutions 
$$
\psi _1({\bf x},t)={\frac 1{\sqrt{V}}}{\frac 1{\sqrt{2E_1}}}\phi _1(s,{\bf p}%
)e^{i{\bf p}\cdot {\bf x}}e^{-iE_1t},\eqno(3.15) 
$$
where $s=1,2$ is the spin index and $\phi _1(s,{\bf p})$ is given by 
$$
\phi _1(s,{\bf p})={\frac 1{\sqrt{1+M_1^2}}}\left( 
\matrix {
u_1(s, {\bf p}) \cr
M_1u_1(s, {\bf p})\cr
}\right) ,\eqno(3.16) 
$$
with 
$$
M_1={\frac{m_\mu -m_e+R}{2\delta }}.\eqno(3.17) 
$$
and $u_1(s,{\bf p})$ is the Dirac spinor 
$$
u_1(s,{\bf p})=\sqrt{E_1+m_1}\left( 
\matrix {
\chi^{(s)} \cr
{{\bf \sigma}\cdot {\bf p}\over E_1+m_1}\chi^{(s)}\cr 
}\right) ,\eqno(3.18a) 
$$
satisfying the Dirac equation $(\gamma ^\mu p_\mu -m_1)u_1(s,{\bf p})=0$,
and $\chi ^{(s)}$ satisfies the normalization condition 
$$
\chi ^{\dagger (s)}\chi ^{(s)}=1.\eqno(3.18b) 
$$
For the other positive energy solution $E_2=\sqrt{p^2+m_2^2}$, we have 
$$
\psi _2({\bf x},t)={\frac 1{\sqrt{V}}}{\frac 1{\sqrt{2E_2}}}\phi _2(s,{\bf p}%
)e^{i{\bf p}\cdot {\bf x}}e^{-iE_2t},\eqno(3.19) 
$$
with $\phi _2(s,{\bf p})$ given by 
$$
\phi _2(s,{\bf p})={\frac 1{\sqrt{1+M_2^2}}}\left( 
\matrix {
u_2(s, {\bf p}) \cr
M_2u_2(s, {\bf p})\cr
}\right) ,\eqno(3.20) 
$$
and $M_2$ defined as 
$$
M_2={\frac{m_\mu -m_e-R}{2\delta }}.\eqno(3.21) 
$$

We notice here that because $M_1M_2=-1$ we can write $\phi _2(s,{\bf %
p})$ in terms of $M_1$ as 
$$
\phi _2(s,{\bf p})={\frac 1{\sqrt{1+M_1^2}}}\left( 
\matrix {
M_1u_2(s, {\bf p}) \cr
-u_2(s, {\bf p})\cr
}\right) ,\eqno(3.22) 
$$
and $u_2(s,{\bf p})$ is the Dirac spinor 
$$
u_2(s,{\bf p})=\sqrt{E_2+m_2}\left( 
\matrix {
\chi^{(s)} \cr
{{\bf \sigma}\cdot {\bf p}\over E_2+m_2}\chi^{(s)}\cr 
}\right) ,\eqno(3.23) 
$$
satisfying the Dirac equation $(\gamma ^\mu p_\mu -m_2)u_2(s,{\bf p})=0$.

Similarly for the solutions of negative energies $-E_{1,2}$ we have the
following eigenfunctions: 
$$
\psi _3({\bf x},t)={\frac 1{\sqrt{V}}}{\frac 1{\sqrt{2E_1}}}\phi _3(s,{\bf p}%
)e^{-i{\bf p}\cdot {\bf x}}e^{+iE_1t},\eqno(3.24) 
$$
with $\phi _3(s,{\bf p})$ given by 
$$
\phi _3(s,{\bf p})={\frac 1{\sqrt{1+M_2^2}}}\left( 
\matrix {-M_2v_1(s, {\bf p}) \cr
v_1(s, {\bf p})\cr
}\right) ={\frac 1{\sqrt{1+M_1^2}}}\left( 
\matrix {v_1(s, {\bf p}) \cr
M_1v_1(s, {\bf p})\cr
}\right) ,\eqno(3.25) 
$$
and 
$$
v_1(s,{\bf p})=\sqrt{E_1+m_1}\left( 
\matrix {
{{\bf \sigma}\cdot {\bf p}\over E_1+m_1}\chi^{(s)}\cr 
\chi^{(s)} \cr
}\right) ,\eqno(3.26) 
$$

$$
\psi _4({\bf x},t)={\frac 1{\sqrt{V}}}{\frac 1{\sqrt{2E_2}}}\phi _4(s,{\bf p}%
)e^{-i{\bf p}\cdot {\bf x}}e^{+iE_2t},\eqno(3.27) 
$$
with $\phi _4(s,{\bf p})$ given by 
$$
\phi _4(s,{\bf p})={\frac 1{\sqrt{1+M_1^2}}}\left( 
\matrix {
M_1v_2(s, {\bf p}) \cr
-v_2(s, {\bf p})\cr
}\right) ,\eqno(3.28) 
$$
and 
$$
v_2(s,{\bf p})=\sqrt{E_2+m_1}\left( 
\matrix {
{{\bf \sigma}\cdot {\bf p}\over E_2+m_2}\chi^{(s)}\cr 
\chi^{(s)} \cr
}\right) .\eqno(3.29) 
$$

\section{Field Quantization, Anti-commutation Relations, and Wave Functions}

We introduce the matrix $U$ 
$$
U=\left( 
\matrix { 
{1 \over \sqrt {1+M_1^2}} & {M_1 \over \sqrt {1+M_1^2}}\cr
{M_1 \over \sqrt {1+M_1^2}} & -{1 \over \sqrt {1+M_1^2}}\cr
}\right) ,\eqno(4.1) 
$$
to make the transformation on the field 
$$
\psi _\nu =\left( \matrix {\psi_e \cr
			\psi_{\mu}\cr }\right) =U\left( \matrix { \phi_1 \cr
		\phi_2 \cr }\right) ,\eqno(4.2) 
$$
which uncouples the interacting Lagrangian given by Eq. (3.1) 
$$
{\cal L}_D={\bar \phi }_1(i\gamma \cdot \partial -m_1)\phi _1+{\bar \phi }%
_2(i\gamma \cdot \partial -m_2)\phi _2.\eqno(4.3) 
$$

Following relations are useful to see that $U$ uncouples Eq. (3.1) 
$$
\delta (M_1+M_2)=m_\mu -m_e;\ \ \ \ R=\delta (M_1-M_2);\ \ \ \ M_1M_2=-1.%
\eqno(4.4) 
$$

Therefore the fields $\phi _1$ and $\phi _2$ describe the ``normal modes''.
To quantize $\psi _\nu $, we expand $\psi _e$ and $\psi _\mu $ in terms of
the normal modes (energy eigenfunctions) found in Sec. 3 
$$
\hat \psi _e({\bf x},t)={\frac 1{\sqrt{V}}}{\frac 1{\sqrt{1+M_1^2}}}\sum_{%
{\bf p}}\sum_s\left [b_1(s,{\bf p}){\frac{u_1(s,{\bf p})}{\sqrt{2E_1}}}%
e^{-iE_1t}+M_1b_2(s,{\bf p}){\frac{u_2(s,{\bf p})}{\sqrt{2E_2}}}%
e^{-iE_2t}\right ]e^{i{\bf p}\cdot {\bf x}}+ 
$$
$$
\left [d_1^{\dagger }(s,{\bf p}){\frac{v_1(s,{\bf p})}{\sqrt{2E_1}}}%
e^{iE_1t}+M_1d_2^{\dagger }(s,{\bf p}){\frac{v_2(s,{\bf p})}{\sqrt{2E_2}}}%
e^{iE_2t}\right ] e^{-i{\bf p}\cdot {\bf x}},\eqno(4.5a) 
$$
$$
\hat \psi _\mu ({\bf x},t)={\frac 1{\sqrt{V}}}{\frac 1{\sqrt{1+M_1^2}}}\sum_{%
{\bf p}}\sum_s\left [M_1b_1(s,{\bf p}){\frac{u_1(s,{\bf p})}{\sqrt{2E_1}}}%
e^{-iE_1t}-b_2(s,{\bf p}){\frac{u_2(s,{\bf p})}{\sqrt{2E_2}}}e^{-iE_2t}\right ]e^{i%
{\bf p}\cdot {\bf x}}+ 
$$
$$
\left [M_1d_1^{\dagger }(s,{\bf p}){\frac{v_1(s,{\bf p})}{\sqrt{2E_1}}}%
e^{iE_1t}-d_2^{\dagger }(s,{\bf p}){\frac{v_2(s,{\bf p})}{\sqrt{2E_2}}}%
e^{iE_2t}\right ]e^{-i{\bf p}\cdot {\bf x}},\eqno(4.5b) 
$$
where the number operators $b_i$ and $d_i$ (i=1,2) satisfy the
anti-commutation relations 
$$
\{b_i(s,{\bf p}),\ b_j^{\dagger }(s^{\prime },{\bf p}^{\prime })\}=\delta
_{ij}\delta _{{\bf p}{\bf p}^{\prime }}\delta _{ss^{\prime }},\eqno(4.6a) 
$$
$$
\{d_i(s,{\bf p}),\ d_j^{\dagger }(s^{\prime },{\bf p}^{\prime })\}=\delta
_{ij}\delta _{{\bf p}{\bf p}^{\prime }}\delta _{ss^{\prime }},\eqno(4.6b) 
$$
and all the other anti-commutators are zero. $b_{1,2}(s,{\bf p})$ annihilates
the normal mode of positive energy $E_{1,2}$ and spin $s$; $d_{1,2}^{\dagger
}$ creates the anti-normal mode of positive energy $E_{1,2}$ and spin $s$.
The vacuum state is defined by 
$$
b_i|0>=d_i|0>=0.\eqno(4.7) 
$$

The total charge operator is 
$$
Q=Q_e+Q_{\mu}=\int{\psi^{\dagger}_{\nu}\psi_{\nu}d^3{\bf x}}= \int{%
(\psi^{\dagger}_e\psi_e+\psi^{\dagger}_{\mu}\psi_{\mu}) d^3{\bf x}}= 
$$
$$
\sum_{{\bf p}}\sum_{s} [b_1^{\dagger}(s, {\bf p})b_1(s, {\bf p})+
b_2^{\dagger}(s, {\bf p})b_2(s, {\bf p})- d_1^{\dagger}(s, {\bf p})d_1(s, 
{\bf p})- d_2^{\dagger}(s, {\bf p})d_2(s, {\bf p})]. \eqno(4.8) 
$$

The following relations hold 
$$
[Q,\ b_i^{\dagger }(s,{\bf p})]=b_i^{\dagger }(s,{\bf p})\ \ \ [Q,\
d_i^{\dagger }(s,{\bf p})]=-d_i^{\dagger }(s,{\bf p}).\eqno(4.9) 
$$
Hence, for a given value of the momentum ${\bf p}$ and spin $s$, there are
four possible normal mode states 
$$
b_1^{\dagger }(s,{\bf p})|0>=|1_{{\bf p}s}>,\ \ \ b_2^{\dagger }(s,{\bf p}%
)|0>=|2_{{\bf p}s}>,\ \ \ \eqno(4.10a) 
$$
$$
d_1^{\dagger }(s,{\bf p})|0>=|-1_{{\bf p}s}>,\ \ \ d_2^{\dagger }(s,{\bf p}%
)|0>=|-2_{{\bf p}s}>.\eqno(4.10b) 
$$

These states differ for the charge ( $\pm 1$, $\pm 2$) 
$$
Q|1_{{\bf p}s}>=|1_{{\bf p}s}>, \ \ \ Q|2_{{\bf p}s}>=|2_{{\bf p}s}>, \ \ \ 
\eqno(4.11a) 
$$
$$
Q|-1_{{\bf p}s}>=-|-1_{{\bf p}s}>, \ \ \ Q|-2_{{\bf p}s}>=-|-2_{{\bf p}s}>,
\ \ \ \eqno(4.11b) 
$$

The above states allow us to construct wave functions in space-time. For
example, the wave function associated with the state $|1_{{\bf p}s}>$ is 
$$
\psi _\nu ({\bf x},t)=\left( 
\matrix {\psi_e({\bf x}, t)\cr
\psi_{\mu}({\bf x}, t)
}\right) =\left( 
\matrix {<0|\hat {\psi}_e({\bf x}, t)|1_{{\bf p}s}> \cr
<0|\hat {\psi}_{\mu}({\bf x}, t)|1_{{\bf p}s}> \cr }\right) = 
$$
$$
=\left( 
\matrix { {1\over \sqrt {1+M^2_1}} \cr
{M_1\over \sqrt {1+M^2_1}}\cr
}\right) {\frac 1{\sqrt{V}}}{\frac 1{\sqrt{2E_1}}}u_1(s,{\bf p})e^{i{\bf p}%
\cdot {\bf x}}e^{-iE_1t}.\eqno(4.12) 
$$
\medskip 

This being a plane wave, gives a stationary probability of finding a
neutrino at a given space-time point. However in any location inside the
volume V there is a probability equal to $({\frac 1{1+M_1^2}})$ of finding
the neutrino in the electron flavor and probability equal to $({\frac{M_1^2}{%
1+M_1^2}})$ of finding it in the muon flavor. Similar considerations can be
applied for the other three states $|2_{{\bf p}s}>$, $|-1_{{\bf p}s}>$, and $%
|-2_{{\bf p}s}>$. Therefore, these states represent states of mixed flavor
at any given space-time point. To obtain states for which we have only one
flavor at a given space-time point we need a superposition of states. A
general state of positive charge, momentum ${\bf p}$ and spin $s$ is given
by 
$$
|\phi _{+}>=[Ab_1^{\dagger }(s,{\bf p})+Bb_2^{\dagger }(s,{\bf p})]|0>,%
\eqno(4.13) 
$$
where $A$ and $B$ specify the amount of each normal mode state of positive
energy present in the state $|\phi _{+}>$.

Similarly, a general state of negative charge, momentum ${\bf p}$ and spin $s
$ is given by 
$$
|\phi _{-}>=[Cd_1^{\dagger }(s,{\bf p})+Dd_2^{\dagger }(s,{\bf p})]|0>,%
\eqno(4.14) 
$$
where $C$ and $D$ specify the amount of each normal mode of negative energy
present in the state $|\phi _{-}>$. The matrix element 
$$
<0|\hat \psi _e({\bf x},t)|\phi _{+}>=\psi _e({\bf x},t)={\frac 1{\sqrt{V}}}{%
\frac 1{\sqrt{1+M_1^2}}}\left [A{\frac{u_1(s,{\bf p})}{\sqrt{2E_1}}}e^{-iE_1t}+M_1B%
{\frac{u_2(s,{\bf p})}{\sqrt{2E_2}}}e^{-iE_2t}\right ]e^{i{\bf p}\cdot {\bf x}},%
\eqno(4.15a) 
$$
gives the probability amplitude of finding a neutrino of momentum ${\bf p}$
and spin $s$ at the space-time point $({\bf x},t)$ with the electron flavor.
In the same way, the matrix element 
$$
<0|\hat \psi _\mu ({\bf x},t)|\phi _{+}>=\psi _\mu ({\bf x},t)={\frac 1{%
\sqrt{V}}}{\frac 1{\sqrt{1+M_1^2}}}\left [M_1A{\frac{u_1(s,{\bf p})}{\sqrt{2E_1}}}%
e^{-iE_1t}-B{\frac{u_2(s,{\bf p})}{\sqrt{2E_2}}}e^{-iE_2t}\right ]e^{i{\bf p}\cdot 
{\bf x}},\eqno(4.15b) 
$$
is the probability amplitude for the muon flavor.

Similarly the matrix elements $<0|\hat \psi _e({\bf x},t)|\phi _{-}>$ and $%
<0|\hat \psi _\mu ({\bf x},t)|\phi _{-}>$ give the probability amplitudes
for finding an electron anti-neutrino flavor and a muon anti-neutrino flavor
respectively.

To take into account the fact that neutrinos (anti-neutrinos) are created
with negative (positive) chiralities, we define the ``observable wave
functions'' as 
$$
\psi _{eL}({\bf x},t)=(1-\gamma _5)\psi _e({\bf x},t),\ \ \ \psi _{\mu L}(%
{\bf x},t)=(1-\gamma _5)\psi _\mu ({\bf x},t) 
\eqno(4.16)
$$
where $\psi _e({\bf x},t)$ and $\psi _\mu ({\bf x},t)$ are given by Eqs.
(4.15).

If ${\bf p}$ is for example along the z-axis, we can choose $\chi ^{(s)}$ as 
$$
\chi ^{(1)}=\left( \matrix {1 \cr 0}\right) ,\ \ \ \ \ \chi ^{(2)}=\left( 
\matrix {0 \cr 1}\right) ,\eqno(4.17) 
$$
and the left-handed Dirac spinor is 
$$
u_L(p,s=2)=(1-\gamma _5)\sqrt{E+m}\left( 
\matrix {
\chi^{(2)} \cr
{ \sigma_z p\over E+m}\chi^{(2)}\cr 
}\right) =\sqrt{E+m}(1+{\frac p{E+m}})\left( 
\matrix {
\chi^{(2)} \cr
-\chi^{(2)}\cr 
}\right) .\eqno(4.18) 
$$
In the limit $E\gg m$, $(1-\gamma _5)u(p,s=1)\simeq 0$.

For ${\bf p}$ along an arbitrary direction ${\bf {\hat n}}$ 
$$
{\bf p}=p[{\hat e}_x sin(\theta)cos(\phi)+{\hat e}_ysin(\theta)sin(\phi)+ {%
\hat e}_zcos(\theta)], \eqno(4.19) 
$$
we can choose the spinors $\chi ^{(s)}$ as 
$$
\chi ^{(1)}=\left( 
\matrix {e^{-{i\over 2}\phi}cos({\theta \over 2})
 \cr e^{{i\over 2}\phi}sin({\theta \over 2 })
}\right) ,\ \ \ \ \ \chi ^{(2)}=\left( 
\matrix {-e^{-{i\over 2}\phi}sin({\theta \over 2})
 \cr e^{{i\over 2}\phi}cos({\theta \over 2 })
}\right) ,\eqno(4.20) 
$$
and the left-handed spinor is still given by Eq. (4.18).

Hence, the observable flavor  neutrino wave functions are 
$$
\psi _{eL}({\bf x},t)={\frac 1{\sqrt{V}}}{\frac{e^{i{\bf p}\cdot {\bf x}}}{%
\sqrt{1+M_1^2}}} [A{\frac{\sqrt{E_1+m_1}}{\sqrt{2E_1}}}(1+{\frac p{E_1+m_1}}%
)e^{-iE_1t}+ 
$$
$$
M_1B{\frac{\sqrt{E_2+m_2}}{\sqrt{2E_2}}}(1+{\frac p{E_2+m_2}}%
)e^{-iE_2t})]\left( \matrix {
\chi^{(2)} \cr
-\chi^{(2)}
}\right) ,\eqno(4.21a) 
$$
$$
\psi _{{\mu }L}({\bf x},t)={\frac 1{\sqrt{V}}}{\frac{e^{i{\bf p}\cdot {\bf x}%
}}{\sqrt{1+M_1^2}}}[ AM_1{\frac{\sqrt{E_1+m_1}}{\sqrt{2E_1}}}(1+{\frac p{%
E_1+m_1}})e^{-iE_1t}+ 
$$
$$
-B{\frac{\sqrt{E_2+m_2}}{\sqrt{2E_2}}}(1+{\frac p{E_2+m_2}}%
)e^{-iE_2t})]\left( \matrix {
\chi^{(2)} \cr
-\chi^{(2)}
}\right) .\eqno(4.21b) 
$$

The coefficients $A$ and $B$ in Eqs. (4.21) are determined through the
initial boundary conditions. Suppose that at $t=0$ 
$$
\psi _{\mu L}({\bf x},t=0)=0,\eqno(4.22) 
$$
we have only the electron flavor present. The other one is 
obtained by the normalization condition 
$$
\int_V{d^3{\bf x}|\psi _{eL}({\bf x},t=0)|^2}=1.\eqno(4.23) 
$$

By imposing the boundary conditions given by Eq. (4.22) and Eq. (4.23) we
obtain the following flavor  wave functions 
$$
\psi _{eL}({\bf x},t)={\frac{e^{i{\bf p}\cdot {\bf x}}}{\sqrt{V}}}{\frac 1{%
1+M_1^2}}{\frac 1{\sqrt{2}}}[e^{-iE_1t}+M_1^2e^{iE_2t}]\left( 
\matrix {
\chi^{(2)} \cr
-\chi^{(2)}
}\right) ,\eqno(4.24a) 
$$
$$
\psi _{{\mu }L}({\bf x},t)={\frac{e^{i{\bf p}\cdot {\bf x}}}{\sqrt{V}}}{%
\frac{M_1}{1+M_1^2}}{\frac 1{\sqrt{2}}}[e^{-iE_1t}-e^{iE_2t}]\left( 
\matrix {
\chi^{(2)} \cr
-\chi^{(2)}
}\right) .\eqno(4.24b) 
$$

Hence the probability densities of finding the electron and muon
neutrino flavor  are given respectively  by 
$$
\rho _e(t)={\frac 1V}\left [1-\left( {\frac{2M_1}{1+M_1^2}}\right) ^2sin^2{\frac{%
(E_2-E_1)t}2}\right ],\eqno(4.25a) 
$$
$$
\rho _\mu (t)={\frac 1V}\left( {\frac{2M_1}{1+M_1^2}}\right) ^2sin^2{\frac{%
(E_2-E_1)t}2}.\eqno(4.25b) 
$$

The coefficient $[2M_1/(1+M_1^2)]^2$ is equivalent to $sin^2(2\theta)$ in
Eqs. (2.10). Therefore the field theory treatment 
reduces to the standard
quantum mechanical treatment described in Sec. 2.

As another example, we consider the case in which the neutrino
eigenfunctions of different masses have different momenta ${\bf p}_1$ and $%
{\bf p_2}$ with 
$$
E_1=\sqrt{{\bf p}_1^2+m_1^2},\ \ \ E_2=\sqrt{{\bf p}_2^2+m_2^2}.\eqno(4.26) 
$$
The neutrino flavor wave functions $\psi_{\nu L}({\bf x},t)$
are therefore 
$$
\psi _{eL}({\bf x},t)={\frac 1{\sqrt{V}}}{\frac{M_1}{1+M_1^2}}{\frac 1{\sqrt{%
2}}}[e^{-iE_1t}e^{i{\bf p}_1\cdot {\bf x}}-e^{iE_2t}e^{i{\bf p}_2\cdot {\bf x%
}}]\left( \matrix {
\chi^{(2)} \cr
-\chi^{(2)}
}\right) ,\eqno(4.27a) 
$$
$$
\psi _{{\mu }L}({\bf x},t)=-{\frac 1{\sqrt{V}}}{\frac 1{1+M_1^2}}{\frac 1{%
\sqrt{2}}}[M_1^2e^{-iE_1t}e^{i{\bf p}_1\cdot {\bf x}}+e^{iE_2t}e^{i{\bf p}%
_2\cdot {\bf x}}]\left( \matrix {
\chi^{(2)} \cr
-\chi^{(2)}
}\right) ,\eqno(4.27b) 
$$
where we have assumed that at a given space-time 
point (${\bf x}={\bf 0}$, $t=0$), we have only the muon flavor.

The probability densities of finding at the space-time point (${\bf x},t$)
 the electron and muon neutrino flavors reduce to Eqs.
(2.10) in the relativistic approximation $|{\bf x}|~\simeq t$.

\section{Conclusions}

We have discussed an explicit model of neutrino flavor mixing in the
framework of quantum field theory. In this model, the equations of motion
for the interacting fields are solved directly and the system is
diagonalized in terms of the two uncoupled free fields $\phi _1$ and $\phi _2$
of mass $m_1$ and $m_2$ respectively. We notice here that because we can
directly diagonalize the Lagrangian we do not need to write the interacting
fields in terms of the free asymptotic fields $\psi _{0e}$, $\psi _{0\mu }$
of mass $m_e$ and $m_\mu $ respectively. We have also derived neutrino 
flavor wave
functions  in such a way that the total flavor
charge is constant. The probability densities, derived from these wave
functions, are in agreement with the standard neutrino oscillation
probabilities, if we take into account the neutrino chirality.

Also, since explicit plane wave solutions for all normal modes have been
obtained, wave packets corresponding to these can be constructed via
standard techniques \cite{Gold}.
$$
$$
\centerline {\bf Acknowledgments}

The author would like to thank Alan H. Guth for his constructive 
criticism which led  to a revision of the paper. She  
would also like to acknowledge fruitful discussions with   
Kenneth Johnson  and the MIT atomic and molecular interferometry group.

\end{document}